\documentclass[pra,onecolumn]{revtex4}
\usepackage{ae} 
\usepackage[T1]{fontenc}
\usepackage[ansinew]{inputenc}
\usepackage{amsmath}
\usepackage{amssymb}
\usepackage{graphicx}

\begin{document}

\title{Entanglement of Formation is Non-monotonic with Concurrence-A simple proof}

\author{Indrani Chattopadhyay}\email{ichattopadhyay@yahoo.co.in}
\affiliation{School Of Science, Netaji Subhas Open University, K-2,
Sec.-V, Salt Lake, Kolkata-700091, India.}
\author{Debasis Sarkar}\email{dsappmath@caluniv.ac.in}
\affiliation{Department of Applied Mathematics, University of
Calcutta, 92, A.P.C. Road, Kolkata-700009, India.\\
Telephone No.- +91 33 23508386, Fax No.- +91 33
23519755.}

\begin{abstract}
In this paper we explore the non-monotonic nature of entanglement of
formation with respect to concurrence for pure bipartite states. For
pure bipartite system, one of the basic physical reason of this
non-monotonicity character is due to the existence of incomparable
states, i.e., the pure bipartite states which are not convertible to
each other by LOCC with certainty.\\
PACS number(s): 03.67.Hk,03.67.Mn, 03.65.Ud.
\end{abstract}

\maketitle

Concurrence is an widely applicable \cite{Cirone} measure for computing entanglement
\cite{wootters1,wootters2} of different physical
systems\cite{majumder}. It is originally proposed to compute
entanglement of formation \cite{purifi,purifi1} for $2\times 2$
systems. For such a system, Entanglement of Formation is proposed to
be a function of concurrence and that function is monotonic in
nature. Afterwards, the measure is generalized for pure as well as
mixed bipartite states in arbitrary dimensions
\cite{Albeverio1,Albeverio2,gour} and also for multi-partite
composite systems \cite{Heydari,Heydari1,Mintert}.

Now, for pure bipartite states, entropy of entanglement is the
unique measure of entanglement \cite{UniqueMeasure}. Consequently,
it is equal to both the entanglement of formation and distillable
entanglement of any pure bipartite state. Although, both
entanglement of formation and distillable entanglement are the most
important measures of entanglement, but only for some special class
of states they are calculable (e.g., \cite{terhal}). So, one has to
probe the relations between some other measures of entanglement or
non-local correlations with the above measures. Concurrence has some
role with entanglement of formation. It is believed that
entanglement of formation of a bipartite state is a monotonic
function of concurrence \cite{Albeverio3}. It is strongly supported
by the evidence of monotonicity character shown for the class of
Isotropic states\cite{I-Concurrence}. It appears to be surprising
that for $2\times 2$ states entanglement of formation is a monotonic
function of concurrence. The generalization of this measure for
higher dimensional systems and the corresponding scheme for
computing entanglement of formation would not suggest us the
non-monotonic character of entanglement of formation with
concurrence. There is no definite physical reason for such
behaviour. Here, we investigate a possible root of this feature with
the existence of incomparable pair of entangled states in pure
bipartite systems. Further, we have found quite similar behaviour of negativity with entanglement
of formation. Thus, our result would also establish the fact that both the concurrence and negativity
have limited scope beyond some lower dimensional bipartite systems and are inadequate measures of entanglement.

The notion of incomparability evolve through the inter conversion of pure bipartite
states under LOCC with certainty \cite{nielsen}. Two
pure bipartite states are said to be comparable if one of them can
be deterministically transformed into the other by
LOCC \cite{nielsen}. To respect the physical restriction of
non-increase of entanglement under LOCC, for a
pair of comparable states $(|\Psi\rangle, ~|\Phi\rangle)$,
$E(|\Psi\rangle) \geq E(|\Phi\rangle).$  But for incomparable pair of pure bipartite states,
the relation is not straight forward. Recently, we have shown
that if a pair of pure bipartite states of same Schmidt rank have the same entropy
of entanglement, then either they are incomparable or locally connected. There exists an
infinite number of mutually incomparable states of same Schmidt rank all having the same
entanglement \cite{IncompareEntropy}. Therefore, entropy of entanglement as the
unique measure for pure bipartite states, fails to express all the
non-local character present within the system. Here, we shall establish
the direct relation between the existence of incomparable states and the non-monotonic
character of entanglement of formation with respect to concurrence. Let us first
consider the following definition of generalized concurrence.

The generalized definition of concurrence for a mixed bipartite state $\rho_{AB}$
in the joint space $H_A\otimes H_B$ of two finite dimensional
Hilbert spaces $H_A$, $H_B$ shared between two parties A, B is defined by,
\begin{equation}
C(\rho_{AB})= \sqrt{2(1-Tr{{\rho_A}^2})}
\end{equation}
where the reduced density matrix $\rho_A$ is obtained by tracing
over the subsystem $B$. Here it is to be noticed that the quantity $Tr{{\rho_A}^2}$ gives a measure of purity of the reduced state $\rho_A$. We assume that a pure $m \times n$ ($m\geq n$) bipartite state has the standard Schmidt form
$|\Psi\rangle_{AB}=\sum_i \sqrt{\mu_i}|a_i\rangle_{A} |b_i\rangle_{B}$, with real valued Schmidt coefficients $\{\mu_i\in [0,1] ~;~~i=1,2,\cdots, k (k\leq \min\{m,n\})\}$ and $\{|a_i\rangle_A\}$, $\{ |b_i\rangle_B\}$ are the orthonormal bases for subsystems $H_A$ and $H_B$ respectively. The integer $k$ is the number of non-zero Schmidt terms of the state $|\Psi\rangle_{AB}$. Then the
concurrence $C(|\Psi\rangle_{AB})$ for this state is given by,
\begin{equation}
C^{2}(|\Psi\rangle_{AB})=4\sum_{i<j} \mu_i \mu_j= 2(1-\sum_{i=1}^k
\mu_i^{2})
\end{equation}  which varies smoothly from
$0$ for pure product states to $2\frac{k-1}{k}$ for maximally
entangled pure states of Schmidt rank $k$. Then, we have the
following result for any pair of comparable pure bipartite states.

\textbf{Theorem:} \emph{For comparable pure bipartite states,
entanglement of formation is  monotone with concurrence.}

Let $|\Psi\rangle, ~|\Phi\rangle$ be any two comparable pure
bipartite states of Schmidt rank $m$ and $n$ respectively, with
Schmidt vectors, $\lambda_{|\Psi\rangle}=(\alpha_1, \alpha_2,
\cdots, \alpha_m)$ and $\lambda_{|\Phi\rangle}= (\beta_1, \beta_2,
\cdots, \beta_n),$ where $\alpha_{i}\geq \alpha_{i+1}\geq 0$ and $\beta_{i}\geq
\beta_{i+1}\geq 0$ and $\sum_{i=1}^{m}
\alpha_{i} = 1 = \sum_{i=1}^{n} \beta_{i}$.
Suppose, $|\Psi\rangle \longrightarrow |\Phi\rangle$ is
possible under deterministic LOCC, then, we must have $m\geq n$ with
$E(|\Psi\rangle) \geq E(|\Phi\rangle)$ and $E(|\Psi\rangle) =
E(|\Phi\rangle)$ if and only if
$\lambda_{|\Psi\rangle}\equiv\lambda_{|\Phi\rangle}$\cite{IncompareEntropy}.
We rewrite the Schmidt vector of $|\Phi\rangle$ as
$\lambda_{|\Phi\rangle}= (\beta_1, \beta_2, \cdots, \beta_m)$ with
$\beta_i=0~~\forall~~i=n+1,\ldots, m$. Now from Nielsen's
criteria\cite{nielsen} for deterministic transformation of pure
bipartite states under LOCC,
$|\Psi\rangle\longrightarrow|\Phi\rangle$ implies
$\lambda_{|\Psi\rangle}\prec \lambda_{|\Phi\rangle}$ (where, $\prec$ is the symbol
for majorization of two real vectors), explicitly,
\begin{equation}
\begin{array}{lcl}
\sum_{i=1}^k \alpha_i \leq \sum_{i=1}^k \beta_i,
~~~~\forall~~k=1,2, \ldots, m-1.
\end{array}
\end{equation}

Then there exists $\{\epsilon_k >0 ~~; ~~k=1,2,\ldots, m-1\}$ such
that
\begin{equation}
\sum_{i=1}^k \beta_i = \sum_{i=1}^k \alpha_i + \epsilon_k,~~
\forall~k=1,2, \ldots, m-1.
\end{equation}
Subtracting each equation from its next one we have
\begin{equation}
\beta_i = \alpha_i + \epsilon_i - \epsilon_{i-1},~~ \forall~k=1,2,
\ldots, m
\end{equation}assuming $\epsilon_0 = 0 = \epsilon_m$.

Comparing the values of concurrences for those two states, we have
\begin{equation}
\begin{array}{lcl}
C^2(|\Psi\rangle)- C^2(|\Phi\rangle)&=& 2(1-\sum_{i=1}^m
{\alpha_i}^2) - 2(1-\sum_{i=1}^m {\beta_i}^2)\\ &=& 2\sum_{i=1}^m
({\beta_i}^2-{\alpha_i}^2)\\ &=& 2\sum_{i=1}^{m}
(\beta_i+\alpha_i)(\beta_i-\alpha_i)\\
&=& 2\sum_{i=1}^{m} (2\alpha_i +
\epsilon_i-\epsilon_{i-1})(\epsilon_i-\epsilon_{i-1})\\
&=& 2\{ 2\sum_{i=1}^{m}\alpha_i (\epsilon_i -\epsilon_{i-1})+ \sum_{i=1}^{m}(\epsilon_i-\epsilon_{i-1})^2\}\\
&=& 2\{2(\sum_{i=1}^{m-1}\alpha_i \epsilon_i - \sum_{i=1}^{m-1}
\alpha_{i+1} \epsilon_i)  \\
& & ~+ \sum_{i=1}^{m}(\epsilon_i-\epsilon_{i-1})^2\} ~ ~ ~ ~ ~ [\because ~ \epsilon_0 = \epsilon_m = 0] \\
&=& 2\{2\sum_{i=1}^{m-1}(\alpha_i  - \alpha_{i+1}) \epsilon_i  + \sum_{i=1}^{m}(\epsilon_i-\epsilon_{i-1})^2\}\\
&\geq& 0 ~~~~ [\because ~~\alpha_i \geq
\alpha_{i+1}~~\forall~~i=1,2, \ldots, m-1].
\end{array}
\end{equation}

Thus for any pair of comparable pure bipartite states we find a
direct relation between entropy of entanglement and concurrence,
i.e., $E(|\Psi\rangle)\geq E(|\Phi\rangle)$ implies
$C(|\Psi\rangle)\geq C(|\Phi\rangle)$. Also, for
comparable states $E(|\Psi\rangle) = E(|\Phi\rangle)$ implies
$\lambda_{|\Psi\rangle}\equiv\lambda_{|\Phi\rangle}$ which further imply,
$C(|\Psi\rangle)= C(|\Phi\rangle)$. Thus for strict relation, $E(|\Psi\rangle) >
E(|\Phi\rangle)$ we have $\lambda_{|\Psi\rangle}\prec
\lambda_{|\Phi\rangle}$ with
$\lambda_{|\Psi\rangle}\not\equiv\lambda_{|\Phi\rangle},$ i.e., for at
least one value of $i$, $\epsilon_i>0$ and ultimately it implies,
$C(|\Psi\rangle)> C(|\Phi\rangle)$.

So, for comparable set of pure bipartite states, entanglement of
formation(which is equal to the entropy of entanglement), is monotone with concurrence. This result is quite
compatible with the case of $2\times 2$ states where entanglement of
formation is always a monotonic function of concurrence and all pure
states of $2\times 2$ system are comparable. From $3\times 3$
composite systems the Hilbert space structure is so much complicated
that one could really understand the full nature of entangled
states. Recently, we found that in the neighborhood of one pure
state of higher Schmidt rank $(\geq 3)$ their exists an infinite
number of other pure states of the same rank which are all
incomparable with the state and all have the same value of entropy
of entanglement \cite{IncompareEntropy}. This feature readily shows
that entanglement of formation is not a monotone function of
concurrence. To explain it, we first consider the following example.\\

\emph{Example:} Consider a pair of pure bipartite
states of $3\times 3$ system represented by the corresponding
Schmidt vectors as,
\begin{equation}
\begin{array}{lcl}
|\Psi_{1}\rangle &\equiv& (.46, .306, .234),\\
|\Phi_{1}\rangle &\equiv& (.43, .3646, .2054)
\end{array}
\end{equation}
The concurrences for these states are,
$C(|\Psi_1\rangle)\simeq1.280016$ and $C(|\Phi_1 \rangle)\simeq 1.279955$. We
compute the value of entanglement of formation for these states by their
entropy of entanglement, which are, $E(|\Psi_1\rangle)\simeq
1.528432837$, $E(|\Phi_1\rangle)\simeq 1.52331025$. Next perturbing
slightly the Schmidt coefficients, we choose another neighboring
pair of pure bipartite states of $3\times 3$ system as
\begin{equation}
\begin{array}{lcl}
|\Psi_{2}\rangle &=& (.43, .3645, .2055),\\
|\Phi_{2}\rangle &=& (.46, .3061, .2339)
\end{array}
\end{equation}
For this new pair the calculated values of concurrences are
$C(|\Psi_2\rangle)\simeq 1.280019$ and $C(|\Phi_2 \rangle)\simeq 1.27998716$.
Whereas the values of the entropy of entanglement for these
states are, $E(|\Psi_2\rangle)\simeq 1.523392983$,
$E(|\Phi_2\rangle)\simeq 1.52839408$.

Both the pairs $(|\Psi_{1}\rangle,~|\Phi_{1}\rangle)$ and
$(|\Psi_{2}\rangle,~|\Phi_{2}\rangle)$ are incomparable in nature.
So, the implicit effect of presence of incomparability is obviously
reflected here. For the first pair of pure bipartite states,
$(|\Psi_{1}\rangle,~|\Phi_{1}\rangle)$ we see
$C(|\Psi_{1}\rangle)>C(|\Phi_{1}\rangle)$ and
$E(|\Psi_{1}\rangle)>E(|\Phi_{1}\rangle)$, while for the second pair
of pure bipartite states $(|\Psi_{2}\rangle,~|\Phi_{2}\rangle)$, we
see $C(|\Psi_{2}\rangle) > C(|\Phi_{2}\rangle)$, but
$E(|\Psi_{2}\rangle) < E(|\Phi_{2}\rangle)$. We could set many such
examples as there are infinite number of pure bipartite states of
same Schmidt rank which are incomparable, having the same value of
entropy of entanglement. Then with a slight perturbation we may
construct infinite number of states which have approximately the
same value of concurrence. For a better understanding we plot the
graphs of concurrences for three equi-entangled classes. By an
equi-entangled class we mean all states of Schmidt rank $3$ having
a specific value of entanglement. Here we present three different
classes having entanglements $1.545$ e-bit, $1.547$ e-bit, $1.550$
e-bit and plotted the graphs of concurrences of those states
against the largest Schmidt coefficients of the states.\\

\begin{center}
FIGURE 
\end{center}

The figure shows that for the states $|\Psi_{A}\rangle,
~|\Phi_{B}\rangle,~|\Psi_{C}\rangle,~|\Psi_{D}\rangle$ represented
by the points A, B, C, D, we have, $E(|\Psi_{A}\rangle) =E(|\Psi_{B}\rangle)
< E(|\Psi_{C}\rangle) = E(|\Psi_{D}\rangle)$ and
$C(|\Psi_{A}\rangle)= C(|\Psi_{D}\rangle)$,
$C(|\Psi_{B}\rangle) = C(|\Psi_{C}\rangle)$. So for the pair of states
$(|\Psi_{A}\rangle,~|\Psi_{C}\rangle)$, we have,
$C(|\Psi_{A}\rangle)> C(|\Psi_{C}\rangle)$ with
$E(|\Psi_{A}\rangle)< E(|\Psi_{C}\rangle)$ and for the pair, $(|\Psi_{B}\rangle,~
|\Psi_{D}\rangle)$,  $C(|\Psi_{D}\rangle)>
C(|\Psi_{B}\rangle)$ with $E(|\Psi_{D}\rangle) >
E(|\Psi_{B}\rangle)$. This does not contrary with the theorem, as
states representing two distinct points(such as the points A and B)
of an equi-entangled class, are necessarily incomparable to each
other.  Here we have plotted only three distinct equi-entangled classes. However, one could draw many
such curves. It shows readily, the entanglement of formation is not a monotonic function of concurrence, even for the pure two-qutrit states. Clearly, this feature establishes the fact that the
entanglement of formation, behaves completely random with the value
of concurrence. This feature of entanglement with concurrence is also observed for other important measures, like, negativity.

Negativity \cite{negativity} is a well known measure of entanglement, which is functionally related with concurrence for $2\times 2$ and $2\times 3$ systems and a lower bound of it. For a general bipartite mixed state $\rho_{AB}$ it corresponds to the absolute value of the sum of negative eigenvalues of $\rho^{T_A}$ (partial transpose of $\rho_{AB}$ with respect to system A) \cite{PPTBE}. It is defined as $N(\rho_{AB})=\frac{\|\rho^{T_A}\|_1-1}{2}$. For the pure bipartite states, $|\Psi\rangle_{AB}=\sum_i \sqrt{\mu_i}|a_i\rangle_{A} |b_i\rangle_{B}$ with Schmidt vector $\lambda_{|\Psi\rangle_{AB}}=(\mu_1, \mu_2,
\cdots, \mu_m)$, Negativity is given by $N(|\Psi\rangle_{AB})=\frac{1}{2}((\sum_i \sqrt{\mu_i})^{2}-1)$. The measure is proposed to quantify the entanglement cost of preparing the state under PPT preserving operations \cite{negPPTcost}. Under numerical study we found, like concurrence it is also an inadequate measure of entanglement and behaves quite similarly with the entropy of entanglement. Considering different classes of equi-entangled states the monotonicity of entropy of entanglement with negativity also does not hold for general pure bipartite states. Thus, in essence, we found, both the negativity and concurrence have limited scope to demonstrate the entanglement behaviour of composite quantum systems.

In conclusion, we have established a physical reason for non-monotonicity
of entanglement of formation with concurrence. Not merely the parametric
behavior of entropy function for higher dimensional systems, but the
incapability of transferring any pair of bipartite pure entangled states
under deterministic LOCC has a fundamental role in such behaviour of two
important measures of entanglement. Our proof is simple and the existence
of so many incomparable states made it much easier to prove. We hope our
result will provide a deep insight on the behaviour of different entanglement
measures in composite quantum systems.

\end{document}